\documentclass[12pt]{article}
\pdfoutput=1 
\usepackage{jheppub}
\usepackage{amsmath}
\usepackage{amssymb}
\usepackage{graphicx}
\usepackage{mathrsfs}
\usepackage[utf8]{inputenc}

\usepackage{subcaption}
\usepackage{float}
\usepackage{afterpage}
\renewcommand{\figurename}{Fig.}

\newcount\xrefpos \xrefpos=0
\newcount\yrefpos \yrefpos=0
\newcount\xput \xput=0
\newcount\yput \yput=0

\def\refpos#1 #2 #3{\global\xrefpos=#1 \global\yrefpos=#2
                         \rlap{$\smash{#3}$}}
\def\put #1 #2 #3{\xput=#1 \yput=#2
                  \advance\xput by -\xrefpos
                  \advance\yput by -\yrefpos
                  \rlap{\kern\the\xput truebp
                        \vbox to 0pt{\vss\hbox{$\displaystyle #3$}
                        \kern\the\yput truebp}}}
\def\beginlabels\refpos#1\endlabels{\hbox{$\refpos#1$}}

\newcommand{\be}{\begin{equation}}
\newcommand{\ee}{\end{equation}}
\newcommand{\bea}{\begin{eqnarray}}
\newcommand{\eea}{\end{eqnarray}}
\def\bse{\begin{subequations}}
\def\ese{\end{subequations}}

\def\IZ{\relax\ifmmode\hbox{Z\kern-.4em Z}\else{Z\kern-.4em Z}\fi}

\def\AdS#1{AdS$_{#1}$}

\definecolor{rust}{rgb}{0.8,0.2,0.2}

\title{Comments on Entanglement Propagation}

\author{Moshe Rozali}
\author{\!, Alexandre Vincart-Emard}
\affiliation{
Department of Physics and Astronomy, University of British Columbia,\\
Vancouver, BC V6T 1Z1, Canada}
\emailAdd{rozali@phas.ubc.ca}
\emailAdd{ave@phas.ubc.ca}
\abstract{We extend our work on entanglement propagation following a local quench in 2+1 dimensional holographic conformal field theories. We find that entanglement propagates along an emergent lightcone, whose speed of propagation $v_E$ seems distinct from other measures of quantum information spreading. We compare the relations we find to information and hydrodynamic velocities in strongly coupled 2+1 dimensional theories. While early-time entanglement velocities corresponding to small entangling regions are numerically close to the butterfly velocity, late-time entanglement velocities for large regions show less regularity. We also generalize and extend our previous results regarding the late-time decay of the entanglement entropy back to its equilibrium value.}

\begin{document}
 
\maketitle

\section{Introduction}
\label{sec:intro}
The generation and propagation of quantum information is a fascinating subject, bringing together insights from quantum information theory, many-body physics and perhaps most surprisingly, studies of the quantum mechanics of black holes. Here we focus on entanglement as a measure of quantum information.

One way to generate entanglement is by quenching the system, i.e. starting the evolution from an atypical excited state of the Hamiltonian, usually generated as the ground state of another, closely related Hamiltonian. The quenching process generates short range entanglement which then evolves and propagates as the system reaches the typical, thermal state\footnote{We note in passing that much of the work on holographic quenches has been done at finite temperature, for example quenching past thermal critical points. Such studies mix quantum entanglement and classical correlations. To directly probe the quantum entanglement of the ground state one needs to work at zero temperature, for example quenching past quantum critical points. While some work in that direction has been done, much more remains to be explored. On the holographic side, the bulk geometry at zero temperature does not involve a regular horizon, which makes both the mechanics and physics quite different from the thermal case.}.

In the holographic context, quenching the system can be achieved by starting at equilibrium and turning on external sources (non-normalizable modes) for marginal or relevant operators, which drive the system out of equilibrium for a finite duration of time. Much attention has been given to global, i.e. spatially homogeneous, quenches. In this case the time-dependence of the entanglement entropy is the observable of interest, and many insights have been gained both in the holographic context, as well as in more traditional approaches to many body physics. Models of entanglement evolution, based on those results, are put forward in \cite{Liu:2013iza,Liu:2013qca,Casini:2015zua,Mezei:2016wfz,Mezei:2016zxg}. It would be interesting to incorporate the spatially-resolved holographic results, discussed here and in \cite{Rangamani:2015agy}, into such models. 

Indeed, the setup of local quenches, whereby the system is excited locally in the spatial domain, provides a spatially-resolved probe of the generation and propagation of entanglement. In \cite{Rangamani:2015agy} we initiated the study of such quenches, and here we continue that study in a more general set of holographic theories involving a charged black hole horizon, corresponding to strongly coupled conformal field theories in 2+1 dimensions at finite charge density. We focus on testing our previous results concerning entanglement propagation in this more general, yet quite similar, context. We are thus able to generalize and improve our original discussion, to test which of our previous results are robust, and to investigate which of our conjectures hold in a more general context\footnote{This is similar in spirit to \cite{Buchel:2015saa,Fuini:2015hba}, where it was found that breaking conformal invariance has only a limited effect on holographic results.}.

Similarly to our previous work, we find that entanglement propagation defines an emergent lightcone structure for the theory. The maximal value of entanglement defines a lightcone, except for narrow transition regimes. We typically find two associated lightcone velocities, one to do with short times, and one with longer times\footnote{Due to numerical limitations, these are not asymptotically long times.}. The associated lightcone velocity $v_E$ in those regimes depends on various parameters, and we have previously found some regularities in the quenches for neutral spacetimes. 

Here we extend that analysis: we find that the early-time velocity seems to be related to the butterfly velocity, while late-time velocities have more complicated phenomenology. We discuss the phenomenology of $v_E$ in this more general setup, and compare our results to other measures of entanglement propagation in that regime. We also discuss the return of the entanglement entropy to its equilibrium value, where we are able to give more precise results than previously due to improved numerics.

The outline of this paper goes as follows: In Section \ref{sec:setupquench} we discuss our setup for local quenches in charged spacetimes, our numerical integration strategy using the characteristic formulation of general relativity, and our holographic calculation of the extremal surfaces encoding the entanglement entropy of regions on the boundary. Section \ref{sec:results} contains analysis of the dynamics of holographic entanglement entropy. We continue our investigation of the emergent lightcone structure that encodes the spatial propagation of entanglement entropy, by including the effects of charge and discussing various mechanisms that may underlie the phenomenology of entanglement dynamics. We also extend our description of entanglement thermalization, for which an improved numerical implementation of the quenches' evolution at late times reveals a logarithmic return to equilibrium rather than an exponential damping. We provide a brief summary of our results in Section \ref{sec:summary} as well as further details on the numerical aspects of this work in Appendix \ref{app:numdetails}.

\vspace{10pt}
\section{Holographic Local Quenches}

In this section we introduce our setup for local quenches in charged spacetimes. The local quench is generated by an inhomogeneous scalar source which is turned on for a finite duration, disrupting the initially uniform energy and charge densities in the process. The resulting bulk solution is found numerically, and the extremal surfaces in that geometry encode the dynamics of the entanglement entropy. Here we describe that setup, before turning to the results in the next section. We focus mostly on differences from \cite{Rangamani:2015agy}, and the reader may wish to consult that reference for additional details.

\label{sec:setupquench}

\vspace{10pt}
\subsection{Scalar Quenches at Finite Density}
\label{sec:ansatz}

We choose our metric to be a generalization of the infalling Eddington-Finkelstein coordinates for black holes in an asymptotically \AdS{4} spacetime \cite{Chesler:2013lia,Balasubramanian:2013yqa}
\begin{align}
 ds^2 = - 2 \,A\, e^{2 \chi} \, dt^2 + 2 \, e^{2 \chi }\, dt \, dr - 2\, F_x \,  dt dx + \Sigma^2 \left( e^{B} \, dx^2 + e^{-B} \,dy^2 \right), 
 \label{ansatz1}
\end{align}
and we introduce a gauge field $V$ in the radial gauge
\be V = V_0 \; dt + V_x \; dx. \ee
The coordinate $r$ denotes the radial bulk coordinate, with the boundary located at $r = \infty$, and $t$ is a null coordinate that coincides with time on the boundary. Our quench, controlled by a relevant scalar on the boundary, will have local support in $x$ while being translationally invariant in the $y$ direction. Hence all the fields under consideration depend only on the coordinates $\{r, t, x\}$ with $\partial_y$ being an isometry. 

This null slicing of spacetime, known as the characteristic formulation, is well adapted to treat gravitational infall problems since the coordinates remain regular everywhere as the quench propagates through the bulk. Our ansatz also provides us with a residual radial diffeomorphism 
\begin{align}
r \rightarrow \overline{r} = r + \lambda(x^\mu) \,,
\label{eq:rrepar}
\end{align}
which we use to fix the coordinate location of the apparent horizon and thus keep the computational domain rectangular. 

The Einstein-Maxwell equations in the presence of a scalar field are given by 
\begin{align} R_{MN} - \frac{R}{2} G_{MN} - \frac{3}{\ell_\text{AdS}^2} G_{MN} =&\; T_{MN}^\Phi + T_{MN}^V, \\ 
\nabla_M F^{MN} =&\; 0  \end{align}
where the matter stress tensors are given by
\begin{align}
T_{MN}^\Phi &= \nabla_M \Phi \nabla_N \Phi + G_{MN} \mathcal{L}_\Phi, \;\;\;\;\; \mathcal{L}_\Phi = -\frac{1}{2} \left( G^{MN} \nabla_M \Phi \nabla_N \Phi + m^2 \Phi^2 \right), \\
T_{MN}^V &=  G^{AB} F_{MA} F_{NB} - \frac{1}{4} F^2 G_{MN}, \;\;\;\;\;\; F = dV.
\end{align}
Before the quench, the spacetime geometry obeys the vacuum Maxwell-Einstein equations and is described by the $\text{RNAdS}_4$ black hole of mass $M$ and charge $Q$
\be ds^2 = - r^2 f(r) \; dt^2 + 2 \; dt \; dr + r^2 \left( dx^2 + dy^2 \right), \;\;\;\;\; f(r) = 1 - \frac{M}{r^3} + \frac{Q^2}{2 r^4}, \ee
and the time-component of the gauge field is
\be V_0 = \mu - \frac{Q}{r}, \;\;\;\;\; \mu \equiv \frac{Q}{r_+}. \ee
The chemical potential $\mu$ is chosen so that $V_0$ vanishes at the event horizon. In fact, RN black holes typically possess two horizons $r_\pm$, which correspond to the two real solutions of $f(r) = 0$. The black hole's Hawking temperature is given by
\be T = \frac{r_+^2 f^\prime(r_+)}{4 \pi}, \ee
and extremality occurs when $T = 0$, i.e. when $Q = \sqrt{3 M r_+/2}$ and the two horizons coincide.

\vspace{10pt}
\subsection{Asymptotic Analysis}
\label{sec:asympt}

We now turn our attention to the asymptotic behaviour of our system. We first make a simplifying choice and take $m^2 \,\ell_\text{AdS}^2= -2$ in order to ensure that the near-boundary expansion of the bulk scalar field is in integer powers of $1/r$ 
\be \Phi(r,t,x) = \frac{\phi_0(t,x)}{r} + \frac{\phi_1(t,x)}{r^2} + \cdots. \ee
Requiring that the Einstein-Maxwell equations in the presence of $\Phi$ are satisfied as $r \rightarrow \infty$ informs us that the gauge field behaves like
\begin{align}
V_0(r,t,x) =&\; \mu(t,x) - \frac{\rho(t,x)}{r} + \cdots \\
V_x(r,t,x) =&\; \mu_x(t,x) + \frac{j_x(t,x)}{r} + \frac{V_x^{(2)}(t,x)}{r^2} \cdots
\end{align}
whereas the metric components have the asymptotic expansion
\begin{align}  
A(r,t,x) =& \;\; \frac{(r+\lambda(t,x))^2}{2} - \partial_t \lambda(t,x) - \frac{1}{4} \phi_0(t,x)^2+ \frac{a^{(3)}(t,x)}{r} + \cdots \\
                     \chi(r,t,x) =& \;\;\frac{c^{(3)}(t,x)}{r^3} + \cdots \\
                     F_x(r,t,x) =& \;\;- \partial_x \lambda(t,x) + \frac{f^{(3)}(t,x)}{r} + \cdots \\
                     \Sigma(r,t,x) =& \;\;r+\lambda(t,x) - \frac{1}{4} \phi_0(t,x)^2 +  \cdots \\
                     B(r,t,x) =& \;\;\frac{b^{(3)}(t,x)}{r^3} + \cdots \,.
\end{align}
The functions $G_{\mu \nu}^{(3)}$ are undetermined by the equations of motion and require the input of boundary data via the stress tensor $T_{\mu \nu}$ \cite{deHaro:2000xn}, defined in its Brown-York form as \cite{Balasubramanian:1999re}
\be T_{\mu \nu} = K_{\mu \nu} - K \gamma_{\mu \nu} + 2\, \gamma_{\mu \nu} - \left( {}^\gamma R_{\mu \nu} - \frac{1}{2} \,{}^\gamma R\, \gamma_{\mu \nu} \right) + \frac{1}{2} \, \gamma_{\mu \nu} \, \phi_0^2, \ee
where $\gamma_{\mu\nu}$ is the induced metric on the boundary, $K_{\mu \nu}, K \equiv \gamma^{\mu\nu} K_{\mu\nu}$ its extrinsic curvatures, and ${}^\gamma R_{\mu\nu}, {}^\gamma R $ its intrinsic curvatures. It is straightforward to show that 
\begin{align} T_{00} =& \;\; 2 a^{(3)} + 4 c^{(3)} + \phi_0 \phi_\text{response}, \\
                    T_{tx} =& \;\; \frac{3}{2} f^{(3)} - \frac{1}{2} \phi_0 \partial_x \phi_0  \,,
\end{align}
and that these components obey the conservation equations
\begin{align} 
\partial_t T_{00} =& \;\; \partial_x T_{tx} + \partial_t \phi_0 \; \phi_\text{response} - (\partial_t \mu_x - \partial_x \mu)^2 - j_x (\partial_x \mu - \partial_t \mu_x),\label{T00dot} \\ 
\partial_t T_{tx} =& \;\; \frac{1}{2} \left( \partial_x T_{00} - 3 \; \partial_x b^{(3)} + \partial_x \phi_0  \; \phi_\text{response} - \phi_0 \; \partial_x \phi_\text{response} \right) + \rho (\partial_x \mu - \partial_t \mu_x). \label{Ttxdot}
\end{align}
In addition to energy and momentum, the electric charge and current are also conserved
\begin{align}
\partial_t \rho =&\; - j_x - \partial_x^2 \mu + \partial_t \partial_x \mu_x, \label{rhodot} \\
\partial_t j_x =&\; V_x^{(2)} + j_x \lambda - \frac{1}{2} \partial_x \rho.
\end{align}

\vspace{10pt}
\subsection{Integration Strategy}
\label{sec:intstrategy}

The characteristic formulation of the Maxwell-Einstein and Klein-Gordon equations conveniently reorganizes the coupled PDEs in two simpler categories: equations for auxiliary fields that are local in time and that obey nested radial ODEs, and equations for dynamical fields that propagate the geometry from one null slice to the next \cite{Chesler:2013lia,Balasubramanian:2013yqa}. Here we outline our numerical integration strategy, and refer the reader to Appendix \ref{app:numdetails} for a discussion on the more technical aspects of our implementation.

We modelled the quench source function as $\phi_0(t,x) =  f(x) g(t)$, with
\begin{align}
f(x) = \frac{\alpha}{2} \left[ \tanh \left( \frac{x + \sigma}{4 s} \right) - \tanh \left( \frac{x - \sigma}{4 s} \right) \right], \;\;\;\;\; g(t) = \text{sech}^2 \left( \frac{t - t_q \Delta}{t_q} \right). 
\label{eq:source}
\end{align}
We let the scalar field profile reach a maximum value $\alpha$ at time $t = t_q \Delta$. We set $t_q = 0.25$ and $\Delta = 8$, and chose the steepness $s$ according to the width $\sigma$ of the perturbation in order to have a smooth profile. By $t=3$, $\phi_0 \approx 0$, and the quench has concluded.

We performed domain decomposition in the radial direction, using 4 domains each discretized by a Chebyshev collocation grid containing 11 points. In doing so, errors located near the boundary or near the apparent horizon remain localized within their respective subdomain \cite{Boyd:2001aa}, thus improving the solutions for auxiliary fields over the entire radial domain. We discretized the spatial direction using a uniformly-spaced Fourier grid over the interval $[-30, 30]$ and used $121$ points for $\sigma = 2$, and $173$ points for $\sigma = 0.5$ to maintain an acceptable spatial resolution as the quench profile propagates further away at later times.

As for the time evolution, we used an explicit fifth-order Runge-Kutta-Fehlberg (RKF) method with adaptive step size to propagate dynamical quantities. Note that we evolved each quench until $t=20$, the approximate time at which the fields perturbations reach the spatial boundaries. We also got rid of high-frequency modes that contaminated our solutions by applying a smooth low-pass filter that discarded the top third of the Fourier modes. However, we remark that it is important not to filter the bulk scalar field $\Phi$ if we want its RKF-propagated boundary profile to agree with the source $\phi_0$ at all times.

\vspace{10pt}
\subsection{Holographic Entanglement Entropy}
\label{sec:HEE}

The next step after obtaining numerical solutions for our local quench is to study the evolution of the holographic entanglement entropy (HEE) of a region $\mathcal{A}$ on the boundary. For simplicity, we consider a strip that extends infinitely in the $y$ direction
\begin{align}
\mathcal{A} = \{(x,y) \,\vert\; x \in (-L,L) , \;\; y \in \mathbb{R} \} \,,\qquad 
\partial\mathcal{A} =\{(x,y) \,\vert\; x = \pm L , \;\; y \in \mathbb{R} \} \,.
\end{align}
The covariant HEE prescription \cite{Hubeny:2007xt} tells us that the entanglement entropy is determined by the area of extremal surfaces anchored on $\partial \mathcal{A}$. It is natural to use the quench's translational invariance to parametrize the extremal surfaces by the coordinates $\tau$ and $y$. The extremal surfaces we are looking for will also be translationally invariant in $y$, and the problem of calculating their area reduces to that of calculating the proper length of the geodesics $X^M(\tau) = \{ t(\tau), r(\tau), x(\tau) \}$ arising from the Lagrangian 
\be \mathcal{L} = G_{yy} \, G_{MN} \dot{X}^M \dot{X}^N. \label{LagGeo} \ee
The resulting system of 3 second order ODEs can be transformed into a system of 6 first order ODEs in the variables
\begin{align} 
\left\{ t, \;\; P_t \equiv \Sigma^2 \, \dot{t}, \;\; r, \;\; P_+ \equiv e^{2 \chi}\left( \dot{r} - A \;\dot{t} \right), \;\; x, \;\; P_x \equiv \Sigma^2 \,\dot{x} - e^{-B} F_x \; \dot{t} \right\},
\end{align}
for which $\mathcal{L} = 2 P_+ P_t + P_x^2$. 

Keeping in mind that the length of a geodesic in an asymptotically AdS spacetime is formally infinite, we introduce a UV cutoff $r = \epsilon^{-1}$ and use a regularization scheme in which we subtract the entanglement entropy of a RNAdS$_4$ geometry expressed with the radial coordinate $\bar{r} = r + \lambda(t,x)$, thus effectively matching asymptotic coordinate charts in both setups and setting $\Delta S_\mathcal{A} = 0$ prior to the quench\footnote{This regularization procedure is equivalent to subtracting the vacuum entanglement entropy for the region $\mathcal{A}$ with a dynamical cutoff $\epsilon_\text{vac}(t,x)$ related to the radial shift $\lambda(t,x)$.}.

To solve the Euler-Lagrange equations derived from (\ref{LagGeo}), we adopt an initial value problem point of view in which the initial conditions at the turning point are
\begin{align} 
\left\{ t = t^*, \;\; P_t =0 , \;\; r = r^*, \;\; P_+ = 0, \;\; x = 0, \;\; P_x = \pm 1 \right\},
\end{align}
and we use a shooting method in $r^*$ so that $x = L$ when $r = \epsilon^{-1}$. Note that the tolerance parameters of the ODE solver must be chosen so that $\mathcal{L} = 1$ along the geodesic, which in turn provides us with a safety check for our solutions.

\vspace{10pt}
\section{Results}
\label{sec:results}

Having described our setup and methods of calculation, we now turn to summarizing the patterns observed in our extended framework. In each case, we provide context by starting our discussion with a brief reminder of our observations for local quenches in neutral spacetimes before broadening the scope of our analysis to account for the effects of charge.
\vspace{10pt}

\subsection{Emergent Lightcone and Entanglement Velocity}
\label{sec:chargeLC}

\vspace{10pt}
\subsection*{Entanglement lightcone}

The local nature of the quenches (having finite energy at infinite volume, i.e. zero energy density) implies that the entanglement entropy of any region $\mathcal{A}$ initially grows with time, reaching a maximum, before inevitably decaying to its pre-quench value as the perturbation dissipates away. Much of our analysis has to do with the spatial structure of that maximum, as a function of the spatial extent $L$ of $\mathcal{A}$ and the time $t$. We find that, except for narrow transition ranges, the curve traced by the maximum in the $L-t$ plane is linear: the spatial propagation of entanglement defines a new lightcone structure, distinct from the causal structure of both bulk and boundary theories.

We note that a similar observation was made in \cite{Nozaki:2013wia}, in which local quenches are implemented as a perturbative approximation to the backreaction caused by a massive infalling particle in pure AdS. In that context, the trajectory traced by $\Delta S_\mathcal{A}(t_\text{max},L)$ in the $L-t$ plane always follows a slope of $v_E = 1$ (additionally, the amplitude of that maximum remains constant throughout). 

It turns out that the structure of our results is much richer since our numerical scheme accounts for the full backreaction of the quench on the geometry. While our data reveals the appearance of an emergent lightcone, this result emerges from the analysis rather than being one of the assumptions put in by hand. Indeed, as we will detail below, we typically find two linear regimes separated by a narrow transition, with distinct velocities at early and late times.

The slope of the curve traced by the maximum, $v_E$, is a natural measure of how fast entanglement propagates spatially. Much of our analysis has to do with analyzing this velocity $v_E$. We find a rich structure in the dependence of the emergent lightcone velocity on parameters. In particular, while it is conceptually similar to other measures of quantum information spreading such as the butterfly or tsunami velocities, we find that it is numerically distinct from them under most circumstances. 

Let us now turn to describing the regularities found in the entanglement velocity $v_E$.

\vspace{10pt}
\subsection*{Entanglement velocity}

As is expected from a relativistic theory, we found that $v_E$ was bounded from above by the speed of light, with the bound being saturated universally in the high temperature regime. 

Perhaps more interesting was the discovery of a lower bound on $v_E$ different from the speed of sound of a three-dimensional CFT, $v_\text{sound} = 1/\sqrt{2} = 0.707$. Indeed, the speed of sound, which underpins the thermalization of energy and momentum on the boundary theory, seemed a likely candidate to track the generation and propagation of entanglement. However, our initial analysis showed that this lower bound lied slightly below $v_\text{sound}$, and in fact was consistently very close to $v_E^{*}(3)$, the tsunami velocity of a Schwarzschild-AdS$_4$ black hole \cite{Liu:2013iza}

\be v_E^{*}(d=3) = \frac{(\eta - 1)^{\frac{1}{2} (\eta - 1)}}{\eta^{\frac{1}{2} \eta}} \Bigg \vert_{d=3} = \frac{\sqrt{3}}{2^\frac{4}{3}} = 0.687, \;\;\; \text{with} \;\; \eta = \frac{2(d-1)}{d}. \label{eq:tsunami} \ee
The tsunami velocity is a holographic measure of how fast entanglement propagates spatially when spacetime is globally quenched and depends uniquely on a black hole's conserved charges. Given the naturalness of this velocity in matters related to entanglement entropy propagation, we conjectured that $v_E$ should be found within the bounds
\be v_E^*(3) = 0.687 \leq v_E \leq 1. \ee
This situation is in a way reminiscent of quantum spin systems, which admit an upper \textit{Lieb-Robinson} bound on the speed at which information can travel despite the absence of relativistic constraints \cite{Lieb:1972wy}. However our holographic calculation also provides us with an unexpected lower bound on information processing based on the properties of spacetime itself.

We now extend our analysis of the structure of the entanglement lightcone and the velocity $v_E$ by including the effects of charge. Our main result persists: in all the cases we examined, the entanglement traces a lightcone structure. We can therefore look more closely at the relation between the entanglement propagation velocity $v_E$ defined by our emergent lightcone structure, and other closely related velocities. We note that while those velocities are conceptually similar, and numerically close to each other for neutral black holes, their dependence on charge is distinct. We can therefore hope to make better distinction between them by examining our results for different parameter ranges, in particular by focusing on the charge dependence.

\vspace{10pt}
\subsection*{Relation to other velocities}

In our simulations for wide quenches we find two stages for entanglement propagation, both exhibiting a lightcone structure, and a narrow transition regime between them. For the early-time results, governing the evolution of small entangling surfaces, it is natural to suspect some relation to the butterfly velocity, quantifying the spatial spread of chaos \cite{Roberts:2014isa,Shenker:2014cwa,Alishahiha:2016cjk}. We note that the presence of charge does not affect its value: $v_\text{butterfly} = \sqrt{3}/2 = 0.866$. 

Indeed, this velocity seems to play a role in our results for the spatial propagation of entanglement entropy: early-time velocities are in the range $v_E \in [0.8, 0.9]$, numerically close to the butterfly velocity. In fact, it was shown that the butterfly velocity naturally characterizes the saturation time for large strip regions in the case of global quenches \cite{Alishahiha:2014cwa}. Since the $L < \sigma$ regime under consideration approximates a global quench for which $t_\text{max}$ can be thought of as the saturation time's counterpart, $v_\text{butterfly}$ seems a likely candidate to quantify the initial spread of quantum information that we observe.

Analytical solutions for global quenches in the limit of very narrow regions $L \ll u_h$ feature two additional characteristic velocities \cite{Kundu:2016cgh}. One of them, dubbed the maximum velocity $v_\text{max} = 0.9464$, yields a measure of the maximum rate of entanglement growth in the quasi-linear regime. However, it is non-causal for $d=2$, and we do not observe $v_E$ in that vicinity in our setup. The other one, called the time-averaged velocity $v_\text{avg} = 0.5991$, corresponds to the instantaneous rate of entanglement growth and approximates the saturation time of $S_\mathcal{A}$ as a function of strip width. Given that we are in an intermediate scaling regime where $L \sim u_h$ and thus receive additional contributions from the bulk, it is not surprising that we did not observe this velocity in our lightcone analysis either.

For the late-time velocities, governing the evolution of larger entangling surfaces, we had previously found a relation to the tsunami velocity, which appears as a lower bound of entanglement propagation in the neutral case. It turns out that the tsunami velocity of RNAdS$_4$ black hole decreases as its charge increases, ultimately vanishing at extremality. If the tsunami velocity serves as a lower bound for all values of the charge, then the addition of charge should change the measured slopes $v_E$ in a predictable way. In particular, we should find that the spatial propagation of the entropy significantly slows down near extremality.

Note however a subtle order of limits issue. Our numerics, performed outside the apparent horizon, are restricted to sufficiently narrow entangling surfaces. This is sufficient for discovering the emergent lightcone structure, which we investigate here. However, the asymptotic IR limit $L\rightarrow \infty$ is a priori distinct and may exhibit different regularities. In particular, even in the extremal limit, the entangling surfaces relevant for the emerging lightcone are not deformed much in the near-horizon region. It may be the case that infinitely wide surfaces are more sensitive to the near-horizon geometry, and thus exhibit a more dramatic behaviour in the near-extremal limit.

\begin{figure*}[t] 
\centering
\begin{subfigure}{0.45\textwidth}
  \includegraphics[width=7cm]{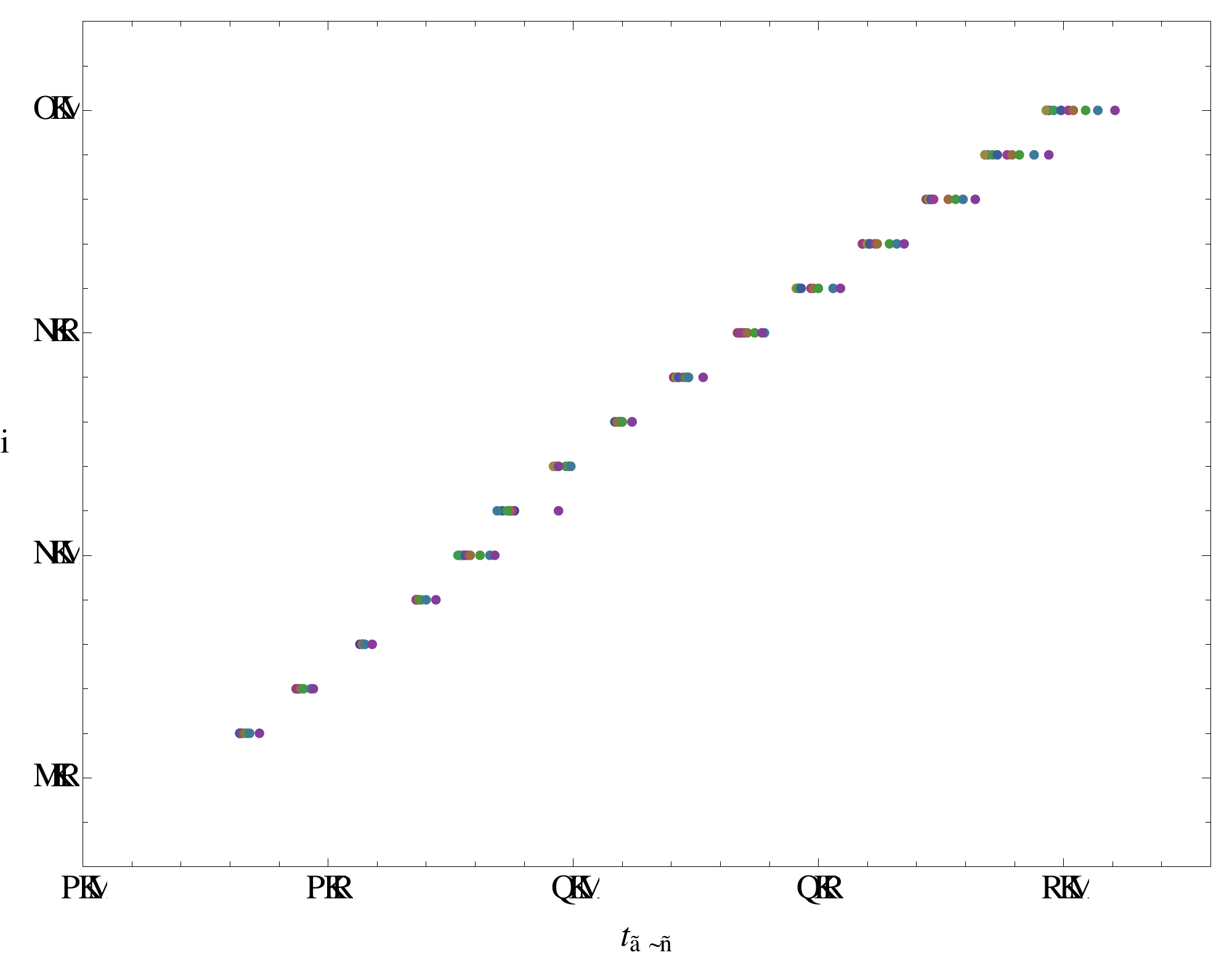}
    \subcaption{$M =0.1,\; \sigma = 2$}
\label{fig_LC1}
\end{subfigure}
\begin{subfigure}{0.45\textwidth}
  \includegraphics[width=7cm]{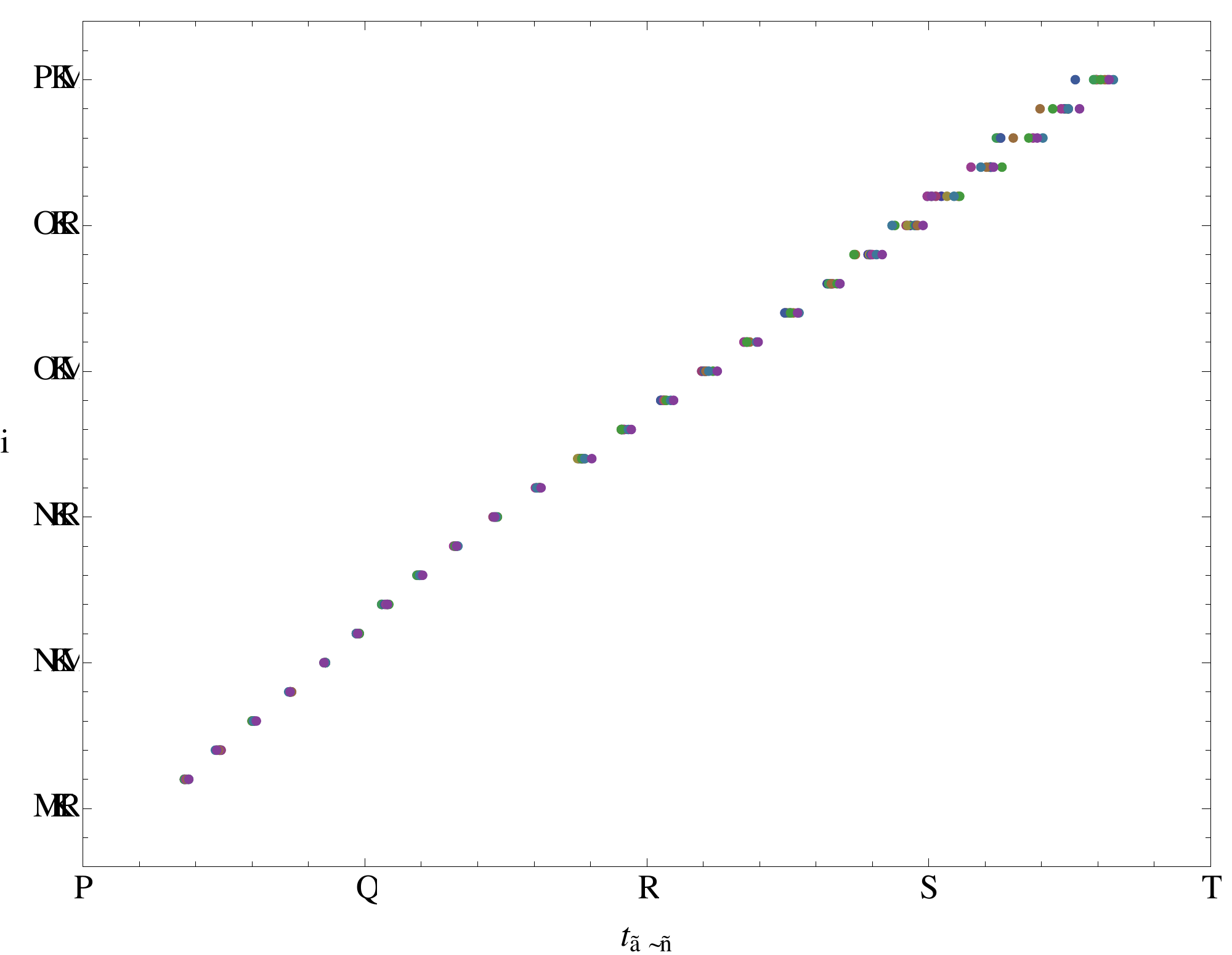}
    \subcaption{$M =0.01,\; \sigma = 2$}
\label{fig_LC2}
\end{subfigure}
\caption{The curves traced by the maximum of $\Delta S_\mathcal{A}(t)$ in the $L-t$ plane. Note that all charged configurations have been included in the same figure to illustrate the weak dependence of the lightcone behaviour with respect to charge. In both cases, the early-time velocities are found in close proximity to the butterfly velocity ($v_E \in [0.8, 0.9]$), whereas the late-time velocities are found within $v_E \in [0.65, 0.71]$, an interval containing various velocities of interest.}
\label{fig_LC}
\end{figure*}

As it turns out, the inclusion of charge does not affect our results in a dramatic way in this regime. Figure \ref{fig_LC} shows the small effect charge has on the lower bound for entanglement propagation speeds; the slopes $v_E$ all fall within the same range for all charged configurations. In the case where the minimal surfaces can penetrate deeper in the bulk, we still observe two linear regimes (as in Fig. \ref{fig_LC2}) corresponding approximately to $L < \sigma$ and $L > \sigma$. The tsunami velocity originally appeared in the large $L$ limit, and we observe that charge only marginally decreases the slope $v_E$.

The range of the lightcone velocities found at large $L$ in our simulations, $v_E \in [0.65, 0.71]$, is also fairly close to other hydrodynamic velocities: $v_\text{sound} = 0.707$ and $v_\text{shear} = 0.665$. The latter is obtained from second-order hydrodynamics results interpreted in terms of the phenomenological Muller-Israel-Stewart theory. This shear velocity, which encodes the velocity of the wavefront of momentum relaxation, is defined as \cite{Baier:2007ix}
\be v_\text{shear} = \sqrt{ \frac{D_\eta}{\tau_\Pi} } \approx 0.665, \ee
where $D_\eta$ is the effective shear ``diffusion" constant obtained from an analysis of the sound pole, and the hydrodynamic parameter $\tau_\Pi$ is the shear relaxation time, which can be calculated from AdS/CFT \cite{Bhattacharyya:2008mz}
\be D_\eta = \frac{1}{4 \pi T}, \;\;\; \text{and} \;\;\; \tau_\Pi = \frac{3}{4 \pi T} \left[ 1 - \frac{1}{2} \left( \log 3 - \frac{\pi}{3 \sqrt{3}} \right)\right]. \ee
As this velocity has to do with entropy production, it can naturally affect the evolution of holographic entanglement entropy in our setup.

In summary, it remains unclear exactly what phenomena come into play to influence entanglement propagation in the late-time regime, where we find an emergent lightcone. On one hand, we have seen that the slope traced by $\Delta S_\mathcal{A}(t_\text{max},L)$ does not decrease as we approach extremality, which suggests that the charged tsunami velocity does not provide an appropriate description of the lower bound for $v_E$. Additionally, our analysis remains inconclusive as to the relevance of the neutral tsunami velocity $v_E^*(3)$. We also see that the entanglement velocity is fairly close to hydrodynamical velocities related to entropy production. As such we are unable to disentangle the various effects which may influence entanglement propagation, and it is entirely possible that different mechanisms may compete to influence the shape of the entanglement lightcone in the late-time regime, resulting in the variations observed in $v_E$.

\vspace{10pt}
\subsection{Entanglement Maximum}

\begin{figure*}[h] 
\centering
\begin{subfigure}{0.45\textwidth}
  \includegraphics[width=7cm]{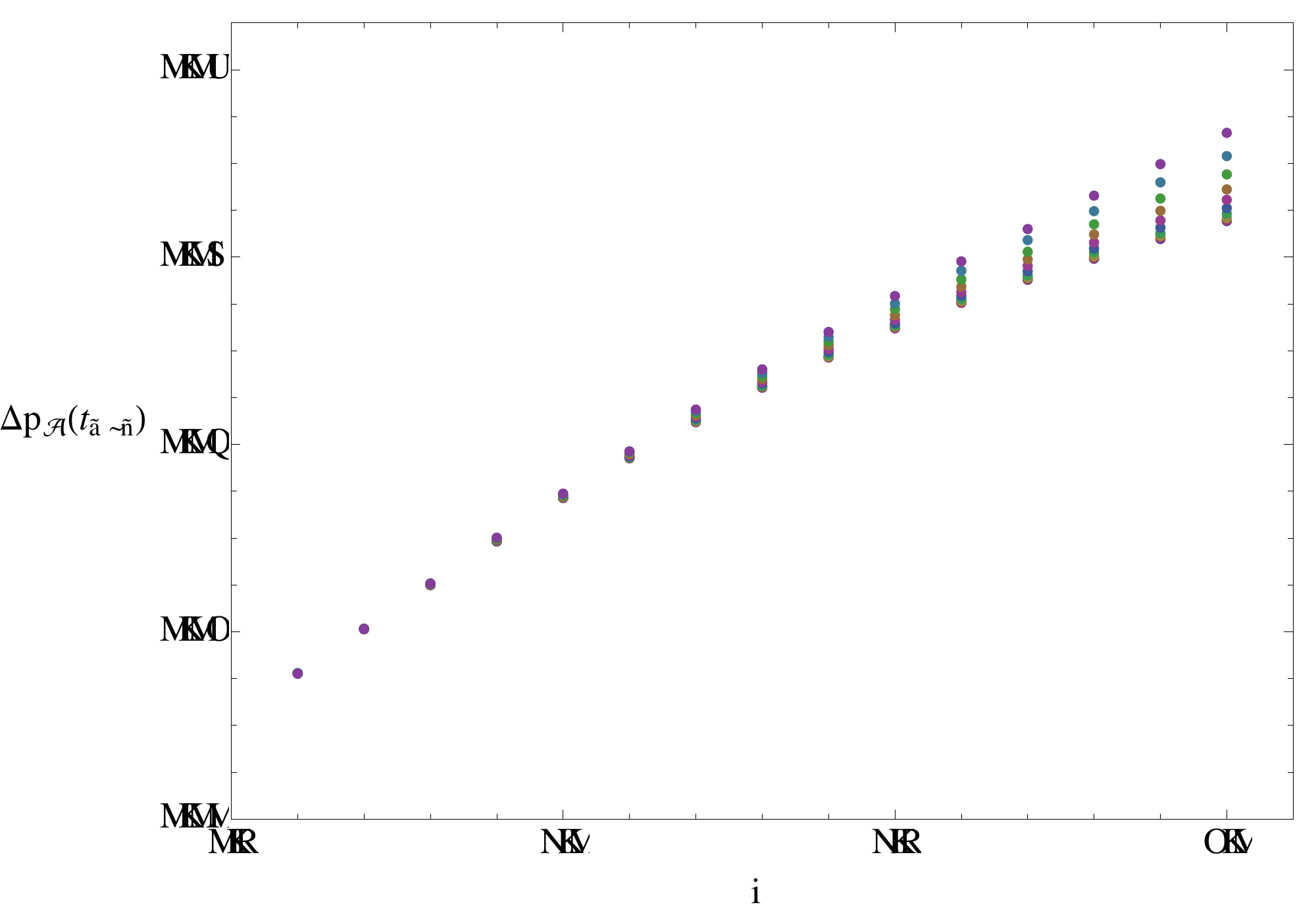}
    \subcaption{$M =0.1,\; \sigma = 2$}
\label{fig_s1}
\end{subfigure}
\begin{subfigure}{0.45\textwidth}
  \includegraphics[width=7cm]{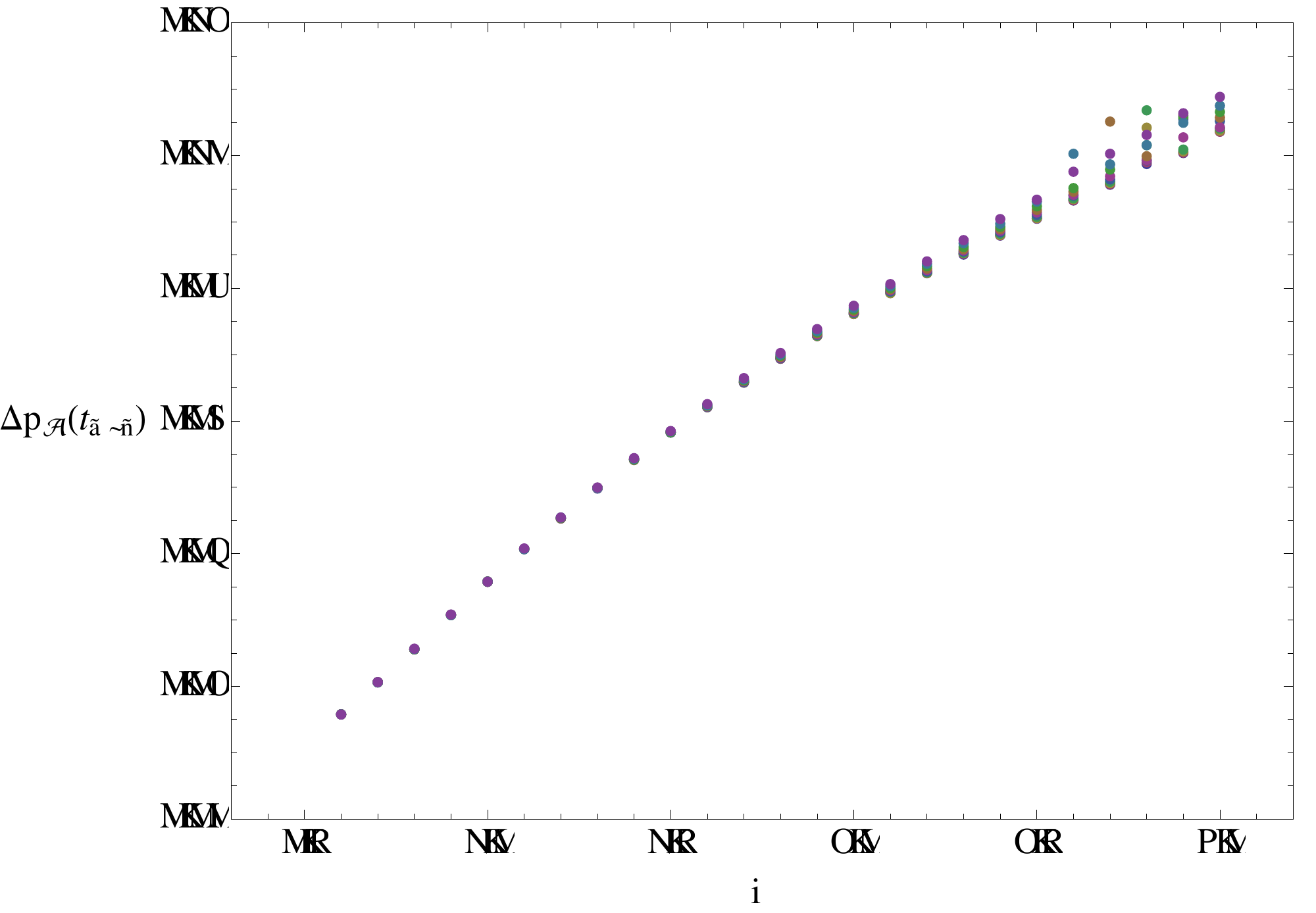}
    \subcaption{$M =0.01,\; \sigma = 2$}
\label{fig_s2}
\end{subfigure}
\caption{The maximum of $\Delta S_\mathcal{A}(t)$ as a function of strip width $L$ for $\alpha = 0.1$. Note that all charged configurations have been included in the same figure to illustrate the weak dependence of the entanglement entropy with respect to charge.}
\label{fig_SL}
\end{figure*}

In the neutral case, we found that the value of the entanglement entropy maximum $\Delta S_\mathcal{A}(t_\text{max})$ increased linearly with the size $L$ of the entangling region for small $L$. This increase was also quantified as a function of the scalar source's maximal amplitude $\alpha$ 
\be \frac{\partial}{\partial L} \Delta S_\mathcal{A}(t_\text{max},L; \alpha) \sim \alpha^2. \ee
For fixed amplitudes, we observe that the maximum $\Delta S_\mathcal{A}(t_\text{max})$ was not affected by the addition of charge for small $L$, and increased marginally when changing $Q$, even as we approach extremality (see Figure \ref{fig_SL}). Thus, our previously discovered regularities seem robust to the addition of charge.

\pagebreak
\subsection{Entanglement Decay}
\label{sec:HEEdecay}

We now turn our attention to the late-time behaviour of holographic entanglement entropy. Our earlier work on local quenches showed evidence that the process of return to equilibrium was best described by an exponential damping
\be \Delta S_\mathcal{A}(t) = a_1 e^{- a_2 (t- a_3)} + a_4. \ee
The parameters $a_i$ depended on the particular features of the quench but did not seem to follow any discernible pattern. However, our analysis was limited by the quality of our numerical quench solutions. In particular, the bulk fields could not be propagated past $t=9$ without loss of accuracy at large $x$ and large memory requirements. We managed to evolve the quenches up until $t=20$ in a reasonable time by making a few modifications, including increasing the spatial resolution by discretizing the $x$ direction with a uniform Fourier grid and by solving the radial ODEs for the auxiliary fields independently for each discretized $x_j$.

\begin{figure}[h!] 
\vspace{20pt}
\centering
    \includegraphics[width=11cm]{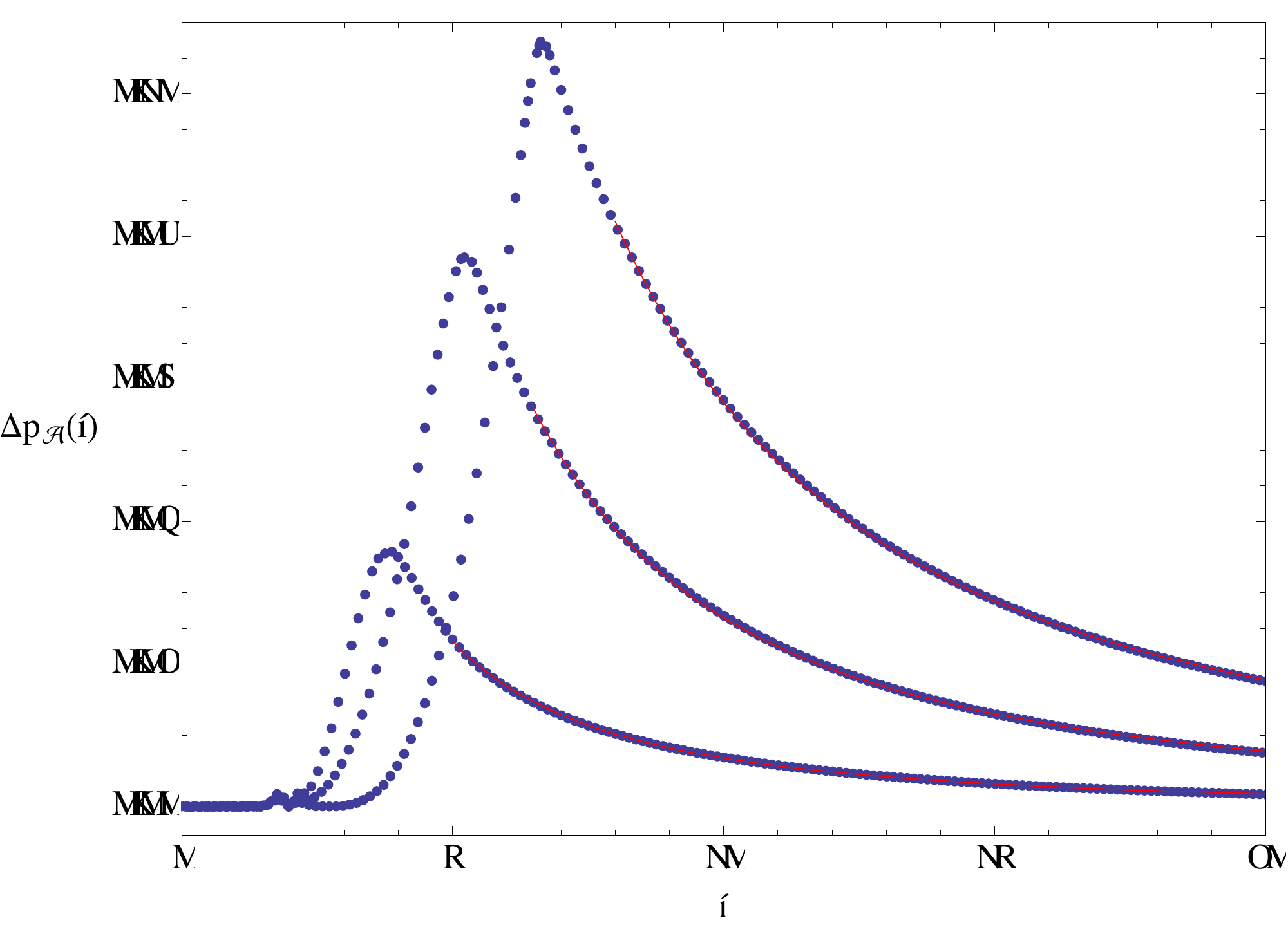}
    \caption[]{The late-time behaviour of the entanglement entropy closely follows the logarithmic decay (\ref{logdecay}) for $L= \{1,2,3\}$, from top to bottom, for $\alpha = 0.1$, $M=0.01$, $\sigma=2$ and $Q=0.04$. In this particular case, the best fit exponents are, respectively, $\delta = \{1.36, 1.48, 1.49\}$. }
\label{fig_HEEdecay}
\end{figure}

These improvements allowed us to investigate the late-time behaviour of the entanglement entropy over much larger time intervals. This additional information revealed that the exponential decay we observed previously was due to fitting the late-time data over too short of a time interval. In fact, the new data instead suggests that 
\be \Delta S_\mathcal{A}(t) \sim \frac{ a_1 \log t + a_2}{t^\delta},\label{logdecay} \ee
is a much better fit, as illustrated in Figure \ref{fig_HEEdecay}. This result is more in line with those derived from spin chain models \cite{Eisler:2007aa}.

\begin{figure*}[h] 
\vspace{10pt}
\centering
\begin{subfigure}{0.45\textwidth}
  \includegraphics[width=7cm]{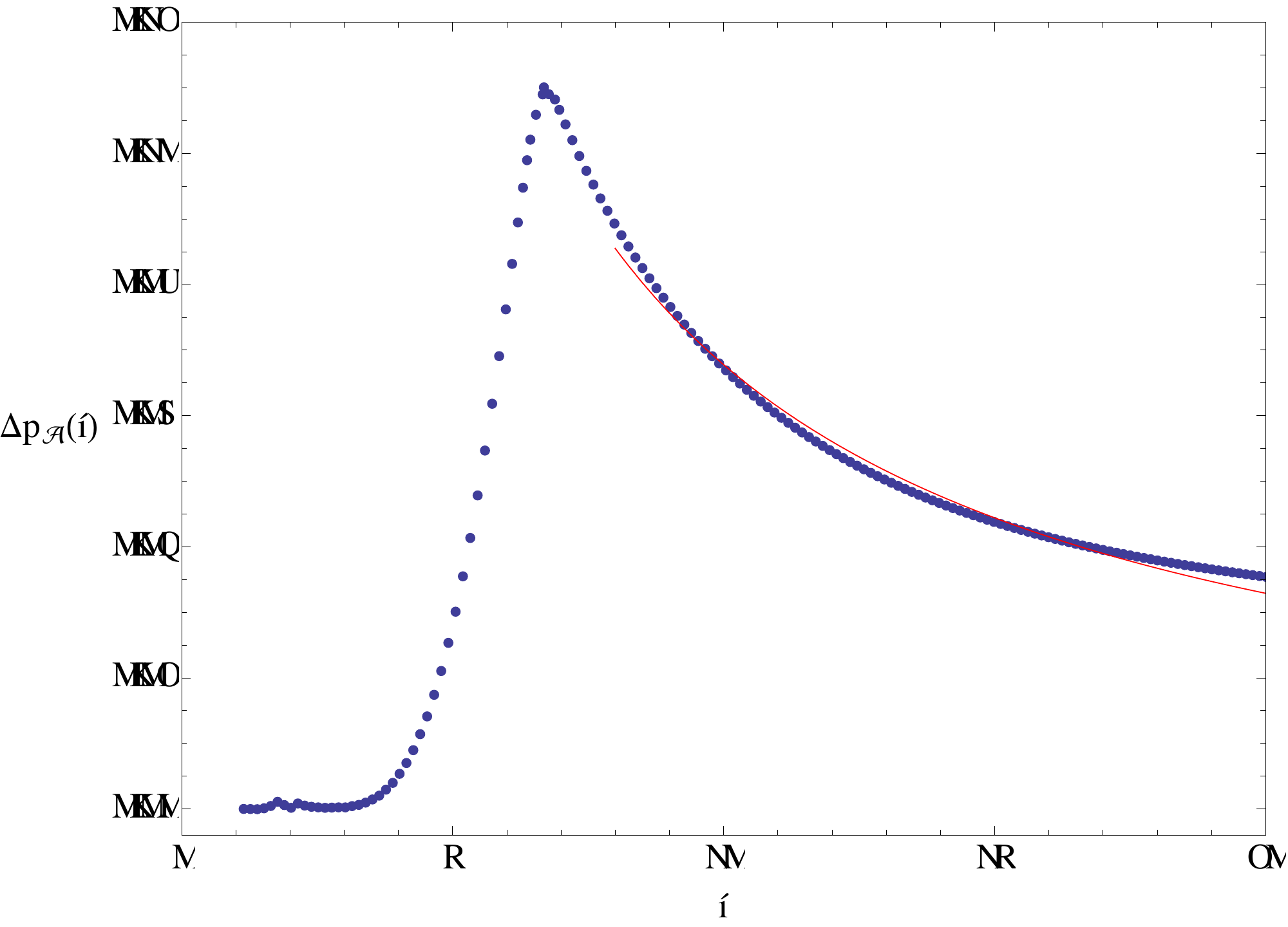}
    \subcaption{$M =0.1,\; \sigma = 2, \; L=3.1$}
\label{fig_S_Ext1}
\end{subfigure}
\begin{subfigure}{0.45\textwidth}
  \includegraphics[width=7cm]{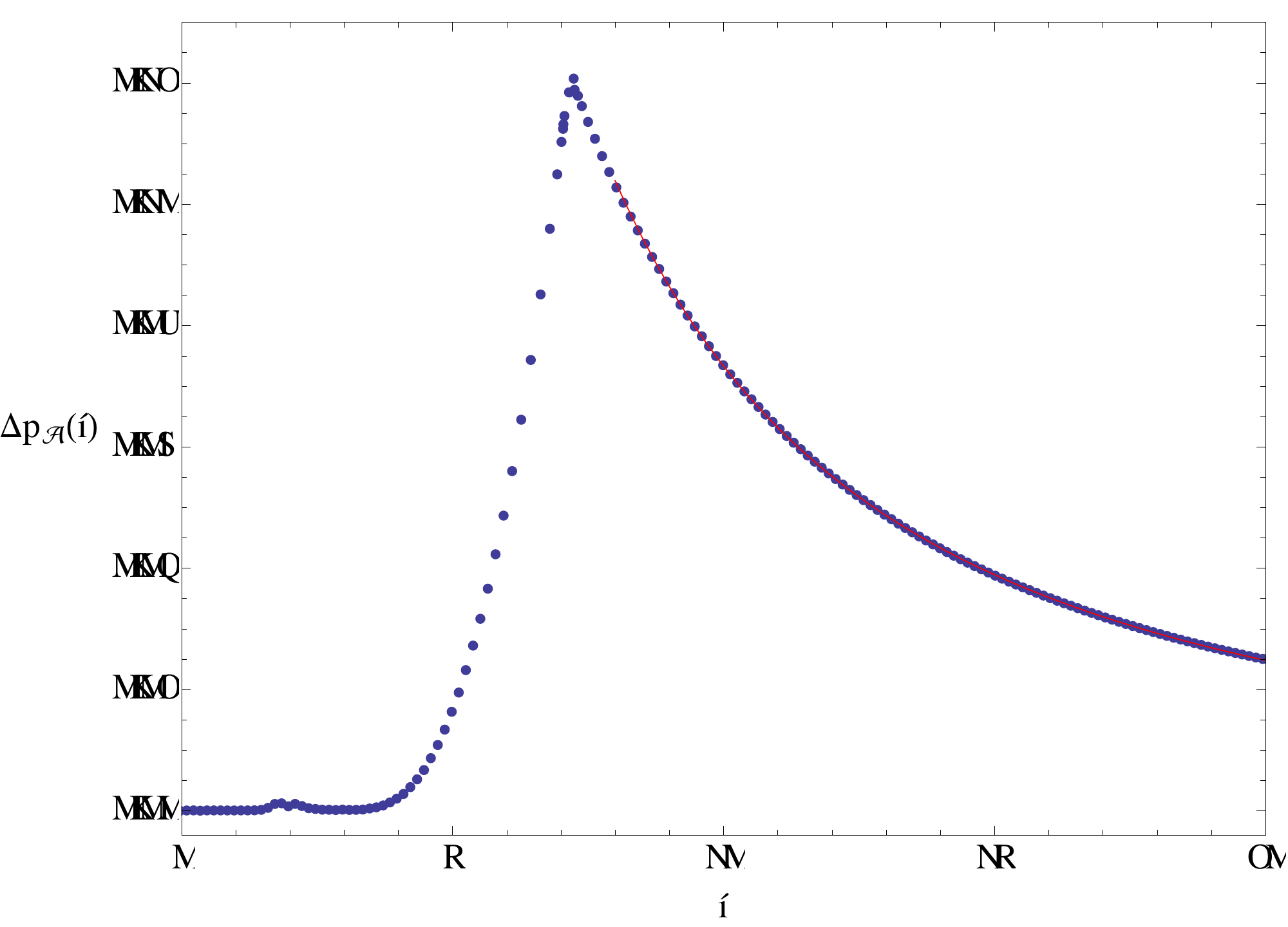}
    \subcaption{$M =0.01,\; \sigma = 2, \; L=3.4$}
\label{fig_S_Ext2}
\end{subfigure}
\caption{The decay of $\Delta S_\mathcal{A}(t)$ and its best logarithmic fit for $Q= 0.99\;Q_\text{ext}$ and $\alpha = 0.1$. The sizes $L$ have been chosen such that the extremal surfaces probe the near-horizon geometry at one point during the quench's evolution, i.e. $L$ is taken as large as the quench allows it to be. We find $\delta = 1.5$ for the figure on the right. }
\label{fig_S_Ext}
\end{figure*}

Interestingly, the best-fit exponents $\delta$, obtained by a least-square fit, are generally clustered around either $\delta = 1$ or $\delta = 1.5$, which marks a departure from the prediction $\Delta S_A(t) \sim t^{-6}$ made in the perturbative analysis of \cite{Nozaki:2013wia}. Our findings show that there is a complex interplay between the size $L$, the initial energy density $M$, the initial charge density $Q$, and the amount of injected energy in the characterization of entanglement entropy's return to equilibrium. When $M=0.1$, the logarithmic decay fits the data with $\delta = 1.5$ at low $Q$ and small $L$ for both $\sigma = 0.5$ and $\sigma=2$. However, (\ref{logdecay}) becomes a bad fit as either the charge and/or the size of $\mathcal{A}$ are increased, as showcased in Figure \ref{fig_S_Ext1}. We find that the breakdown occurs around $Q \sim Q_\text{ext}/2$. 

\begin{figure}[t!] 
\vspace{10pt}
\centering
    \includegraphics[width=9cm]{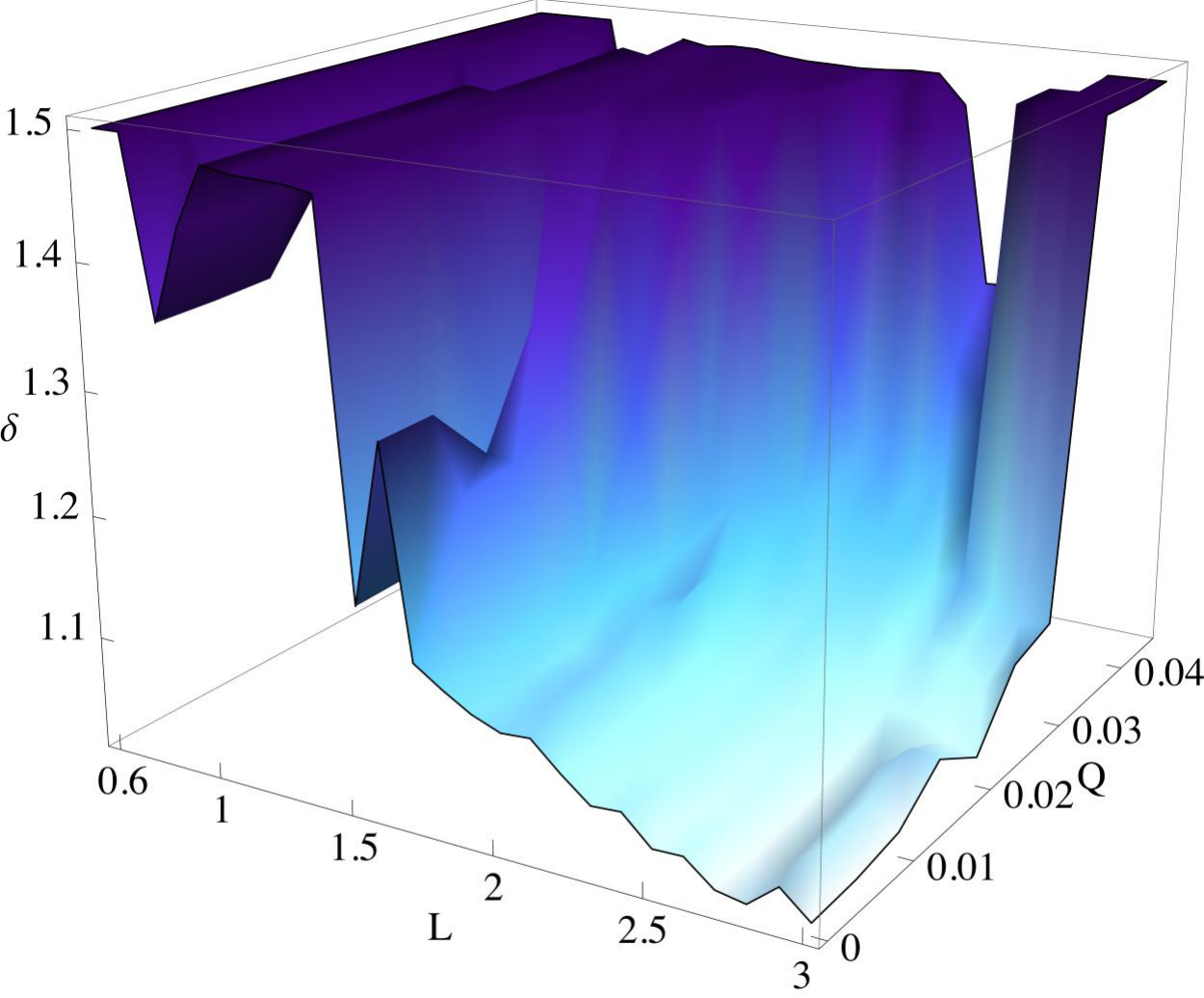}
    \caption[]{This figure illustrates the sharp transition between $\delta = 1$ and $\delta = 1.5$ in the logarithmic decay of the HEE as a function of $Q$ and $L$ for $M=0.01$ and $\sigma = 2$.}
\label{fig_HEEdecay_delta}
\end{figure}


In contrast, the logarithmic return to equilibrium fits the data for all values of $Q$ and $L$ when $M=0.01$. When $\sigma = 0.5$, thermalization is dominated by $\delta = 1$ except for near-extremal black holes $Q = 0.99\; Q_\text{ext}$, for which $\delta = 1.5$ no matter the size of the entangling surface. Taking $\sigma = 2$ reveals an even richer picture in which we observe a sharp transition between decays characterized by $\delta = 1$ and $\delta = 1.5$. As illustrated in Figure \ref{fig_HEEdecay_delta}, the late-time evolution of holographic entanglement entropy in the neutral, large size limit is fitted best with $\delta = 1$. The exponent $\delta = 1.5$ appears either as extremality is approached, as in the $\sigma = 0.5$ case, or in the small $L$ limit, as in the $M=0.1$ case.

These observations lead us to believe that the late-time behaviour of $\Delta S_\mathcal{A}(t)$ is influenced not only by the parameters characterizing the geodesics and the geometry of the unquenched spacetime, but also by the amount of energy injected by the scalar. As such it is hard to disentangle and generalize our findings when the underlying competing processes arbor inherently different length scales.

\vspace{20pt}
\section{Summary}
\label{sec:summary}

We have studied the spatial propagation of entanglement entropy following a local excitation of the system. We find that the entanglement generically propagates along an emergent lightcone, whose velocity may change over a narrow transition regime. In our simulation we find early and late-time velocities, and look at their dependence on parameters and relation to other interesting information and hydrodynamical velocities.

The early-time entanglement velocity for small strips seems similar to the butterfly velocity. As both have to do with the initial propagation of quantum information, we find that relation plausible, especially as it mirrors an analytical result derived in an analogous global quench scenario. We are however unable to disentangle the various effects that could influence the late-time entanglement velocity: the propagation in that regime seems likely to be controlled by a combination of many mechanisms.

We are also able to exhibit some universality in the logarithmic return of the entanglement to its equilibrium value. In particular, the relation to known results for spin chains in 1+1 dimensional CFTs is intriguing.

There are very few avenues to investigate the propagation of quantum information in higher-dimensional strongly coupled conformal field theories. We hope that the phenomenology we present can illuminate that difficult subject: in particular it would be instructive to have a simple model incorporating the regularities we find in the holographic results. We hope to return to these issues in the future.  

\vspace{20pt}
\section*{Acknowledgements}

We thank Mukund Rangamani for initial collaboration on the paper and many insightful comments on related subjects. We thank Hong Liu and Mark Mezei for interesting conversations. The research is supported by a Discovery grant from NSERC.

\clearpage
\appendix

\section{Numerical Details}
\label{app:numdetails}

The characteristic formulation of Einstein's equations in the presence of matter reorganizes all the fields into two categories: auxiliary fields obeying radial ODEs that can be solved sequentially, and dynamical fields which are used to evolve the geometry from one null slice to the next. This separation of fields can be achieved by expressing time derivatives in terms of the directional derivative along outgoing null geodesics, $d_+ = \partial_t + A \; \partial_r$, thereby completely eliminating the presence of $A$ from our scheme. Changing to a compact variable $u = 1/r$, we rewrite the fields appearing in our equations as
\begin{equation}  
\begin{split}
\Phi(u,t,x) \equiv& \;\; \phi(u,t,x) u, \\
E_r(u,t,x) \equiv& \;\; e_r(u,t,x) u^2, \\
\Sigma(u,t,x) \equiv& \;\; \frac{1 + \lambda(t,x) u}{u}  - \frac{1}{4} \phi(u,t,x)^2  u,  \\
B(u,t,x) \equiv& \;\; b(u,t,x) u^2, \\
\chi(u,t,x) \equiv& \;\; c(u,t,x) u^2, \\
 F_x(u,t,x) \equiv& \;\;- \partial_x \lambda(t,x) + f_x(u,t,x), \\
 d_+ \Sigma(u,t,x) \equiv& \;\; \frac{(1 + \lambda(t,x)u )^2}{2 u^2} + \tilde{\Sigma}(u,t,x), \\
 d_+ \Phi(u,t,x) \equiv& \;\; - \frac{1}{2} \phi(u,t,x) + \left( \tilde{\Phi}(u,t,x) + \frac{1}{2} \partial_u \phi(u,t,x) \right), \\
 d_+ B(u,t,x) \equiv& \;\;  \tilde{B}(u,t,x) u^2, \\
 A(u,t,x) \equiv& \;\; \frac{(1+\lambda(t,x) u)^2}{2 u^2} + a(u,t,x) ,
\end{split}
\end{equation}
in order to subtract the divergent parts. The field $E_r(r,t,x)$ above is defined as
\be E_r = \partial_r V_0 + \frac{e^{-B}}{\Sigma^2} F_x \; \partial_r V_x \sim F^{tr} \ee
in order to decouple the radial equations satisfied by $V_0$ and $F_x$. However we note that the equations for $d_+ B$ and $d_+ V_x$ form a linear system of radial ODEs that cannot be decoupled.

Given initial conditions specified by $\phi$, $\lambda$, $b$ and $V_x$ all being 0, as well as the CFT data $T_{00}$ and $T_{tx}$, we can solve the radial ODEs for the auxiliary fields $c$, $e_r$, $f_x$, $V_0$, $\tilde{\Sigma}$, $\tilde{\Phi}$, and for the coupled system $\tilde{B}$ and $d_+ V_x$, in that order. These fields obey the boundary conditions 
\begin{align}
\partial_u c(u=0) =&\; - \frac{1}{12} \lambda \phi_0^2 + \frac{1}{6} \phi_0 \phi_1, \\
e_r(u=0) =&\; \rho, \\
f_x(u=0) =&\; 0 \;\;\; \text{and} \;\;\; \partial_u f_x(u=0) = f^{(3)} = \frac{2}{3} T_{tx} + \frac{1}{3} \phi_0 \partial_x \phi_0, \\
V_0(u=0) =&\; \mu, \\
\tilde{\Phi} (u=0) =&\; - \phi_1 - \lambda \phi_0 + \partial_t \phi_0, \\
\tilde{B}(u=0) =&\;  \frac{1}{6}\left(  (\partial_x \phi_0)^2 - \frac{1}{2} \phi_0 \partial_x^2 \phi_0 - \partial_x T_{tx} \right) - \frac{1}{2} j_x\left( \partial_t \mu_x - \partial_x \mu - \frac{1}{2} j_x \right) - \frac{1}{2} \partial_u^2 b \Big\vert_{u=0}, \\
d_+ V_x(u=0) =&\; \partial_t \mu_x - \frac{1}{2} j_x.
\end{align}
There are two options when treating the field $\tilde{\Sigma}$, one of which is to impose the condition
\begin{align}
 \left[ d_+\Sigma - \frac{e^{-B}}{2\,\Sigma} \left( F_x \,\partial_x B - \partial_x F_x - e^{-2 \chi} \,F_x^2 \,
 \frac{\partial_r \Sigma}{\Sigma} \right) \right]_{r=r_h} = 0, \,
\label{appcond} 
\end{align}
which determines the location of the apparent horizon as the boundary of trapped surfaces \cite{Rangamani:2015agy}. Our second option is to set 
\be \partial_u \tilde{\Sigma}(u=0) = \frac{1}{2} T_{00} - \frac{1}{3} \phi_0 \phi_1 - \frac{1}{12} \lambda \phi_0^2 \ee
on the boundary, as required by self-consistency of the equations of motion. Either conditions imply the other; imposing the latter should yield the former and vice-versa, and we can use this as a safety check for our numerics.

Now that we have solved for the necessary auxiliary fields, we have to propagate the solutions along null slices. In order to propagate $\lambda$, we require a horizon stationarity condition, obtained by differentiating (\ref{appcond}) with respect to time and thus ensuring that the location of the apparent horizon remains fixed at all times. This procedure yields a boundary value problem in $x$ for the field $A(u_h,t,x)$. We can then extract $\partial_t \lambda$ from the relation
\be d_+ \Sigma = \partial_t \lambda + A - d_+ \left( \frac{1}{4} \phi^2 \right) \ee
evaluated at the horizon. The same equation in turn enables us to solve for $A$ everywhere in the bulk since $\lambda$ does not depend on the radial coordinate. With $A$ in hand, it now becomes straightforward to extract the time derivatives for $b$, $\phi$ and $V_x$ from the solutions for $d_+B$, $d_+ \Phi$ and $d_+ V_x$, and from the definition of $d_+ = \partial_t + A \; \partial_r$. At this point, all that is left to do is propagate these fields, along with $T_{00}$, $T_{tx}$ and $\rho$ using the conservation equations (\ref{T00dot}), (\ref{Ttxdot}), and (\ref{rhodot}), and to repeat the process on new slices until satisfied with the time evolution of the quench. 

\vspace{10pt}
 \bibliographystyle{jhep}
 \bibliography{quench}

\end{document}